\documentclass[aps]{revtex4}
\usepackage{color}
\usepackage{psfrag}
\usepackage{dcolumn}
\usepackage{bm}
\usepackage{float}
\usepackage[latin1]{inputenc}
\usepackage[spanish,english]{babel}
\usepackage{amsfonts}
\usepackage{amssymb,epsf}
\usepackage{graphicx}
\usepackage{epstopdf}
\usepackage{amsmath,amssymb}
\usepackage{pdflscape}
\usepackage{hyperref}

\begin{document}

\title{\textbf{Anisotropic} magnetized neutron star}
\author{ Gholam Hossein Bordbar$^{1,2}$%
\footnote{
email address: ghbordbar@shirazu.ac.ir} and Mohammad Karami$^{1}$\footnote{%
email address: mohammad.karami68@gmail.com}}
\affiliation{$^{1}$ Physics Department and Biruni Observatory, Shiraz University, Shiraz 71454, Iran\\
$^{2}$ Department of Physics and Astronomy, University of Waterloo, 200
University Avenue West, Waterloo, Ontario, N2L 3G1, Canada}

\begin{abstract}
As we know, the effect of strong magnetic field causes the anisotropy for the magnetized compact objects.
Therefore, in this paper, we have studied the structure properties of anisotropic case of magnetized neutron
star. We have derived the equation of state (EoS) of neutron star matter
for two forms of magnetic fields, one uniform and one density dependent.
We have solved the generelized Tolman-Oppenheimer-Volkoff equations to examine the maximum mass and corresponding
radius, Schwarzschild radius, gravitational redshift, Kretschmann scalar, and Buchdahl theorem
for this system. It was shown that the maximum mass and
radius of neutron star are increasing functions of the magnetic field. Also redshift, strength of
gravity, and Kretschmann scalar increase as the magnetic field increases. In addition, the dynamical stability of
anisotrop neutron star has been investigated, and finally a comparison with the empirical results
has been made.
\end{abstract}

\maketitle

\section{Introduction}

Neutron stars are very dense compact objects with the strongest known
magnetic fields in the world \cite{Camenzind}. The surface magnetic field
strength for ordinary neutron stars is about $10^{12}-10^{15}G$, while for
magnetars, it is much higher than $10^{15}G$. Despite such estimations for the strength of magnetic field,
its structure is not well known for us. Internal magnetic fields for these stars are estimated about $10^{18}G$ \cite{Reisenegger,Woltjer,Lai}.
The fact that strong magnetic fields are presented in the most compact
astrophysical objects, and they have significant implications for several stellar
properties has motivated to study the equation of state (EoS) of
magnetized neutron stars, both without considering \cite%
{Ferrer,Canuto,Broderick,Fayazbakhsh,Abrahams,Suh,Perez} and considering
\cite{Broderick2,Mao,Felipe,Yue,Dexheimer,Dong,Casali} the magnetic-field
interaction with the particle anomalous-magnetic-moment. An
important feature of EoS in a strong uniform magnetic field is that the pressure is  anisotropic \cite%
{Ferrer,Canuto}. It is proposed that the magnetic field arises naturally in
neutron stars as a consequence of thermal effects occurring in their outer
crusts. The heat flux through the crust, which is carried mainly by
degenerate electrons, can give rise to a possible thermoelectric instability
in the solid crust which causes horizontal magnetic field components to grow
exponentially with time \cite{Blandford,Pons}. Possible origins of the
magnetic fields of neutron stars include the fossil field hypothesis, also
called flux conservation, field generated by a dynamo process in the
progenitor star, and the thermomagnetic effect in the neutron stars crust.
When the magnetic fields in a neutron star are strong, the effects of the
magnetic field cause anisotropy in the components of the energy-momentum
tensor, creating two components of pressure. In the presence of a uniform
magnetic field, both  matter and the field contributions to the space
like components of the energy-momentum tensor become anisotropic. The degree
of pressure anisotropy increases as the magnitude of the magnetic field
increases \cite{Sinha}. Any extended object that is placed in an external field
will experience different forces throughout its extent and the result is a
tidal deformation \cite{Chatzi}. Historically, significant efforts have been
made to gain a comprehensive understanding of the properties of anisotropiy
effects, with the hope of producing suitable models of compact stars. This
work was first mentioned by Lematre \cite{Lematre} in the structure and
evolution of compact objects. However, the interest in studying the
distribution of anisotropic relative matter in general relativity has been
revived by Bowers and Liang \cite{Bowers}. They set up and solved the
equations of hydrostatic equilibrium for a locally anisotropic, static, and
spherically symmetric distribution of matter. They found a change in maximum
mass $M$ and surface redshift $z$. Specifically to solve the mathematical problem of
developing anisotropic fluid sphere models for a coupled system of
three independent nonlinear partial differential equations in five geometric
and dynamic variables namely metric potentials $({\nu })$ and $({\lambda })$
and density $({\rho })$, radial pressure $(p_{r})$ and tangential pressure $%
(p_{t})$. Because the system is poorly defined, field equations can be
solved for any metric. This approach is not necessarily fruitful because all
control of the physics of the problem is lost. For example, there are no
equations of state, and this is often seen as  standard for perfect liquids. It
is notable that the difference between the principal pressures is known as
the anisotropic parameter denoted by $({\Delta })$ \cite{Roupas}.
\textbf{Recently, the assumption of pressure anisotropy for the compact objects has been also discussed by Herrera \cite{herrera2020}.
In that paper, It has been shown that
starting from an initially isotropic fluid, at the time scale
under consideration, the evolution leads to an anisotropic
fluid. In fact, Herrera has indicated that the dissipative fluxes, and/or energy density inhomogeneities
and/or the appearance of shear in the fluid flow, force any
initially isotropic configuration to abandon such a condition, generating
anisotropy in the pressure. This means that an initial fluid configuration with isotropic
pressure would tend to develop pressure anisotropy as it evolves, under
conditions expected in stellar evolution.}

In our previous works we have studied the structure properties of neutron stars in the absence and presence of magnetic field where we have considered the isotropic case of neutron star. We used the modern equation of state
(EoS) and calculate the neutron star structure \cite{Bordbar1}. We have computed the structure of cold and hot neutron stars with the quark core and compared them with those of neutron stars without the quark core \cite{Bordbar2, Bordbar7}. We have also studied the effect of cosmological constant and different gravity theories of Einstein-$\Lambda$, $(3+1)$-dimensional rainbow gravity and spin-$2$ massive gravitons, on the properties of neutron star\cite%
{Bordbar3,Bordbar4,Bordbar5,Bordbar6}.
After that we have evaluated the properties of the cold neutron star due to in the presence of a quark core \cite{Bordbar7}.
The properties of spin
polarized neutron matter in the presence of strong magnetic fields at zero
\cite{Bordbar8,Bordbar9}, and finite temperatures \cite{Bordbar10} have been investigated.
As we have mentioned in the above discussions, the strong magnetic field causes the anisotropy for the star.
Therefore, in the present work, we consider the anisotropic case of magnetized neutron star which contains pure neutron
matter to evaluate its structure properties. For this purpose we
calculate the energy of neutron matter by the lowest order constrained variational (LOCV) method in the presence of
magnetic field, and use it to obtain the EoS for this star \cite{Bordbar8,Bordbar9,Bordbar10}.
Finally we solve the generelized Tolman-Oppenheimer-Volkoff equations to calculate the gravitational mass, radius and some of other properties of this system.

\section{Equation of state of anisotrop magnetized neutron star}
We consider a homogeneous and pure system included $N$ interacting particles
with $N^{(+)}$ spin up neutrons and $N^{(-)}$ spin down neutrons
under the influence of the two types of the magnetic field in which one of
them is a uniform magnetic field,
\begin{equation}
\boldsymbol{B}=B{\hat{k}},
\end{equation}
where, $B$ is constant from the center to the surface of the neutron
star. The other form of magnetic field is considered as a function of the
density. For this form, we use a Gaussian function as follows \cite%
{Bandyopadhyay},
\begin{equation}
\boldsymbol{B(\rho )}=B_{surf}+B_{0}[1-\exp (-{\beta (\frac{\rho }{\rho _{0}}%
)^{\theta} )}],
\end{equation}%
where $B_{surf}$ is the magnetic field of the surface of neutron star
that we consider $10^{13}G$ and $B_{0}$ is the interior magnetic field that
is expected for star. Also $\beta $ and $\theta $ are the parameters that
define the magnetic field changes based on the density of neutron star \cite%
{Casali}. These parameters are selected as the magnetic field decrease fast or
slow from the center to the surface of the star. In our case we consider $%
\beta=0.05 $ and $\theta=2 $. The number densities of spin-up and spin-down
neutrons are denoted by ${\rho }^{(+)}$ and ${\rho }^{(-)}$, respectively.
Spin polarization parameter, $\delta $, define as%
\begin{equation}
\delta =\frac{{\rho }^{(+)}-{\rho }^{(-)}}{\rho },
\end{equation}%
where $-1\leq \delta \leq 1$ and ${\rho }$ is the number density of
neutrons in the whole system. The magnetization density of the system is obtained by
\begin{equation}
m={\mu _{n}\delta \rho },
\end{equation}%
where ${\mu _{n}}$ is the magnetic moment of the neutron. Magnetization of a
certain volume of a system {is} determined by integrals
\begin{equation}
M=\int mdV.
\end{equation}

In order to calculate the energy of this system, we use LOCV method as
follows.
We consider a trial many-body wave function in the following form
\begin{equation}
\psi =\digamma \phi ,
\end{equation}%
where $\phi $ is the ground-state wave function of $N$
noninteracting neutrons, and $\digamma $ is a proper N-body correlation
function. Using Jastrow approximation \cite{jastrow}, $\digamma $ can be
replaced by
\begin{equation}
\digamma =S\prod_{i>j}f(ij),
\end{equation}%
where $S$ is a symmetrizing operator. We consider a cluster expansion of the
energy functional up to the two-body term
\begin{equation}
E([f])=\frac{1}{N}\frac{\langle {\psi }|H|{\psi }\rangle }{\langle {\psi }|{%
\psi }\rangle }=E_{1}+E_{2}.
\end{equation}
First, we should calculate the one-body term and two-body term of energy and
then consider the case in which the spin polarized neutron matter is under
the influence of a strong magnetic field. The one-body term $E_{1}$ for spin
polarized neutron matter is given by
\begin{equation}
E_{1}=\sum_{i=+,-}\frac{3}{5}\frac{\hbar ^{2}k_{f}^{{(i)}^{2}}}{2m}\frac{%
\rho ^{(i)}}{\rho },
\end{equation}%
where $k_{f}^{(i)}={(6\pi ^{2}\rho ^{(i)})}^{(\frac{1}{3})}$ is the Fermi
momentum of a neutron with spin projection $i$. The two-body energy $E_{2}$
is
\begin{equation}
E_{2}=\frac{1}{2N}\sum_{ij}\langle {ij}|\nu {(12)}|{ij-ji}\rangle ,
\end{equation}%
where
\begin{equation}
\nu {(12)}=-\frac{\hbar ^{2}}{2m}[f(12),[\bigtriangledown
_{12}^{2},f(12)]]+f(12)V(12)f(12).
\end{equation}
In the above equation, $f(12)$ and $\nu (12)$ are the two-body correlation
function and nuclear potential, respectively. By minimization of the two-body energy with respect to the correlation function,
 we get a set of differential equations. By solving these differential equations, we
can compute the energy of this strongly interacting system (see Refs. \cite{Bordbar8,Bordbar11} for
more details).

Now we consider the case in which the spin polarized neutron matter is under
the influence of a strong magnetic field. Taking the uniform magnetic field
along the $z$ direction, $B=B\hat{k}$, the spin up and down particles
correspond to parallel and antiparallel spins with respect to the magnetic
field. Therefore, the contribution of magnetic energy of the neutron matter
is
\begin{equation}
E_{M}=-M_{z}B,
\end{equation}%
where $M_{z}$ is the magnetization of the neutron matter which is given by
\begin{equation}
M_{z}=N\mu _{n}\delta .
\end{equation}
In the above equation, $\mu _{n}=-1.9130427$ is the neutron magnetic moment
(in units of the nuclear magneton). Consequently, the energy per particle up
to the two-body term in the presence of magnetic field can be written as
\begin{equation}
E([f])=E_{1}+E_{2}-\mu _{n}B\delta .
\end{equation}
%

\begin{figure}[]
\center{\includegraphics[width=6cm]{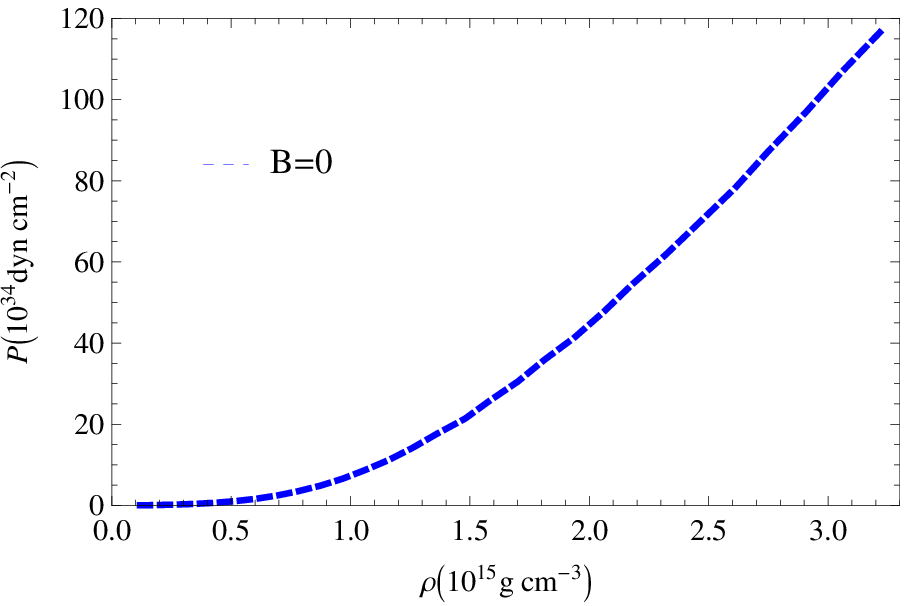}
\includegraphics[width=6cm]
		{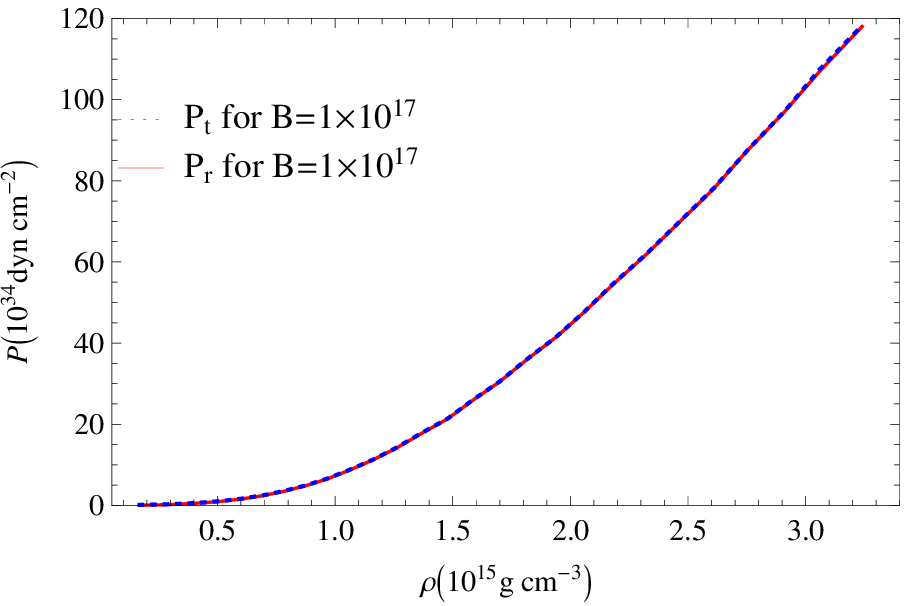}}
\caption{{\protect\small The equation of state of magnetized neutron matter
for $B=0$ (left panel) and (right panel) the radial pressure and tangential
pressure .vs density of magnetized neutron star for $B=1\times10^{17}G$.}}
\label{P0}
\end{figure}

\begin{figure}[]
\center{\includegraphics[width=6cm]{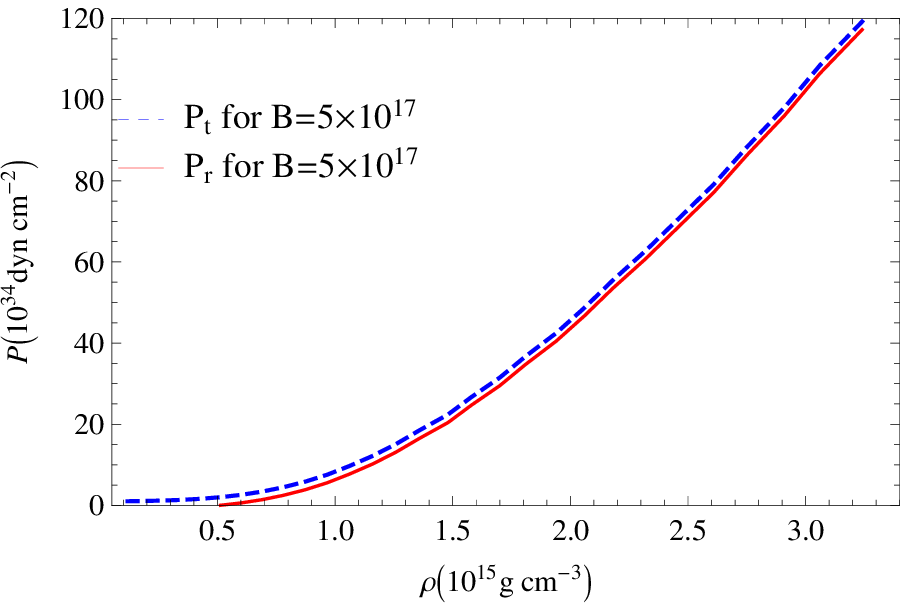}
\includegraphics[width=6cm]{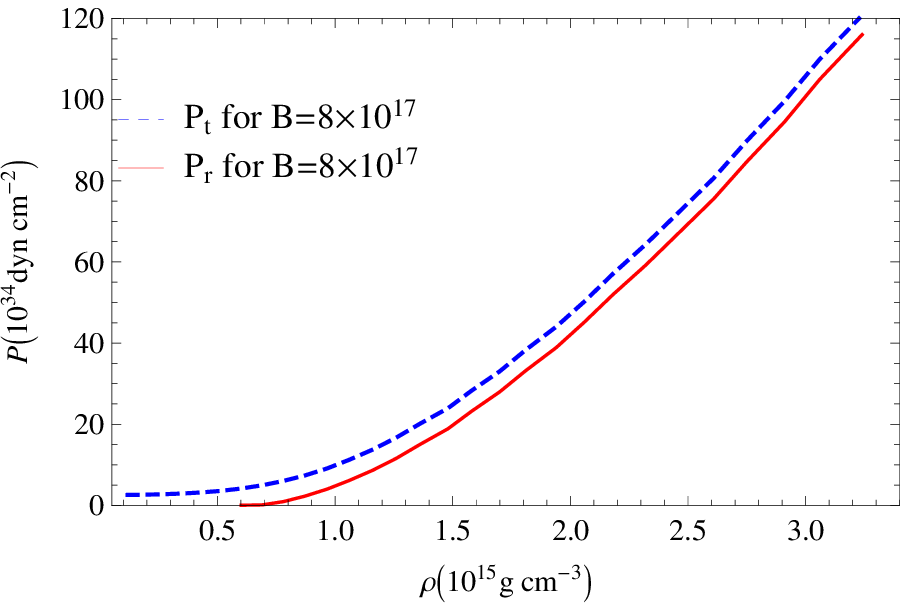}}
\caption{{\protect\small The radial pressure and tangential pressure .vs
density of magnetized neutron star for $B=5\times10^{17}G$ (left panel) and
for $B=8\times10^{17}G$ (right panel). }}
\label{P5}
\end{figure}

\begin{figure}[]
\center{\includegraphics[width=8cm]{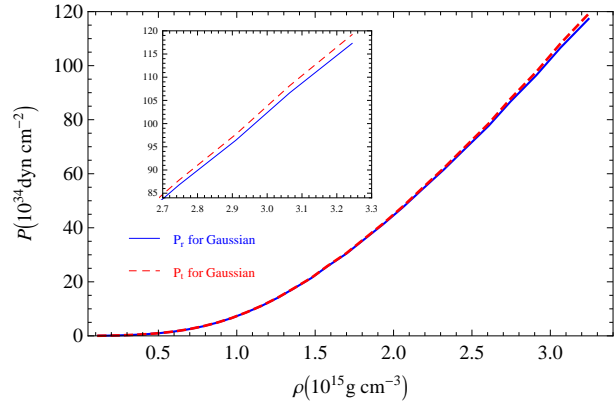}}
\caption{{\protect\small The radial pressure and tangential pressure .vs
density of magnetized neutron star for Gaussian magnetic field. }}
\label{PG}
\end{figure}

\begin{figure}[]
\center{\includegraphics[width=8cm]
		{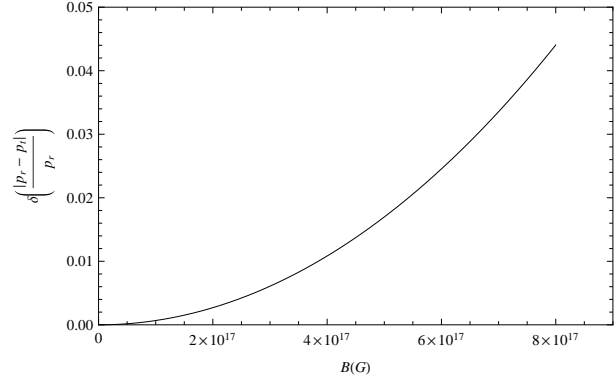}}
\caption{{\protect\small The relative anisotropy parameter $\protect\delta$
vs. magnetic field. }}
\label{Delta}
\end{figure}

From the energy of neutron matter, at each magnetic
field, we can evaluate the corresponding pressure $(P)$ using the following
relation,
\begin{equation}
P(\rho ,B)=\rho ^{2}\biggl(\frac{\partial E(\rho ,B)}{\partial \rho }\biggr)%
_{B},  \label{press}
\end{equation}%
which leads to the equation of state of the system.
%
%
In the strong magnetic field the pressure of neutron
star becomes anisotropic where it has two components \cite{Mallick},
the tangential and radial pressure as follows
\begin{equation}
P_{t}=\rho ^{2}\biggl(\frac{\partial E(\rho ,B)}{\partial \rho }\biggr)_{B}+%
\frac{B^{2}}{8\pi },
\end{equation}%
\begin{equation}
P_{r}=\rho ^{2}\biggl(\frac{\partial E(\rho ,B)}{\partial \rho }\biggr)_{B}-%
\frac{B^{2}}{8\pi }.
\end{equation}
Now, we define the anisotropy parameter as $\Delta =P_{t}-P_{r}$, this
parameter shows the order of anisotropy in the system, where it is the
difference between tangential pressure and radial pressure. As we said
before, we make our calculations for different magnetic fields and compute
the EoS of system for each case of magnetic field. For $B=0$, there is no anisotropy and $P_t=P_r=P$.
 For $B=1\times 10^{17}G$
as one can see in Figure \ref{P0}, there is no significant difference
between tangential pressure and radial pressure, but by increasing the magnetic field we can see that the
 anisotropy increases. For $B=5\times 10^{17}G$ and $B=8\times 10^{17}G$, as one can see
in Figure \ref{P5}, there is a difference between the radial pressure and
the tangential pressure. We have also plotted  density  for the Gaussian
magnetic field in Figure \ref{PG}. It shows that as the density increases, the
difference between the radial pressure and the tangential pressure
increases. Indeed, the anisotropy parameter increases with increasing
density. Here the relative anisotropy parameter can be defined as $\delta =\frac{|P_{r}-P_{t}|}{%
P_{r}}$. Our results in the Figure \ref{Delta} indicate
that $\delta $ increases by increasing the magnetic field.

\begin{figure}[]
\center{\includegraphics[width=8cm]{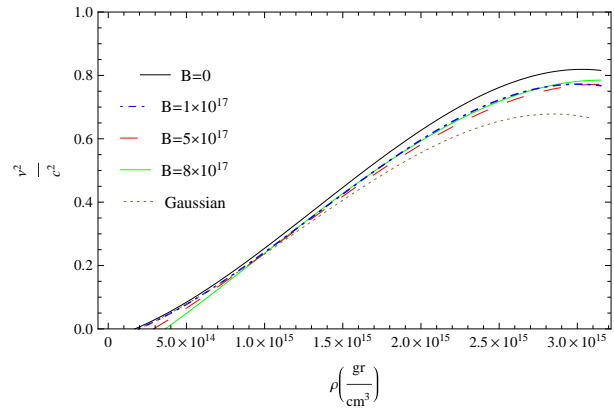}}
\caption{{\protect\small Sound speed vs. density }}
\label{velocity}
\end{figure}

One of the interesting constraints on the EoS is related to the causality condition. In other
words, our calculated EoS should satisfy the condition of causality where the obtained speed
of sound $\left( \nu =\sqrt{\frac{dp}{d\rho }}%
\right) $ should be lower than the speed of light in vacuum. We present our results for the
sound speed versus density in Figure \ref{velocity}. It is evident that for
all magnetic fields, our EoS of the magnetized neutron star matter satisfies
the condition $0\leqslant \nu ^{2}\leqslant c^{2}$ (see Figure \ref%
{velocity}, for more details). So, our EoS of the magnetized neutron star
matter is suitable to study the structure properties of magnetized neutron stars \cite%
{Tews}.
%
%

\section{Structure of anisotrop magnetized neutron star}
In astrophysics, Tolman-Oppenheimer-Volkoff (TOV) equations expresses the
structure of an object with a spherical symmetry that is in hydrostatic
equilibrium \cite{TVO1,TVO2,TVO3},
\begin{equation}
\frac{dP}{dr}=-\frac{Gm(r)\rho (r)}{r^{2}}\left( 1+\frac{P(r)}{\rho
(r)c^{2}}\right) \left( 1+\frac{4\pi r^{3}P(r)}{m(r)c^{2}}\right) \left( 1-%
\frac{2Gm(r)}{rc^{2}}\right) ^{-1},  \label{tvo1}
\end{equation}%
where $\rho (r)$ is the energy density, $G$ is the gravitational constant
and
\begin{equation}
m(r)=\int_{0}^{r}4\pi r^{\prime 2}\rho (r^{\prime })dr^{\prime },
\label{tov2}
\end{equation}%
gives the gravitational mass inside a radius $r$. For the anisotropic case
we use the generalized TOV equation as follows \cite{Riazi,Ponce},
\begin{equation}
\frac{dP_{r}}{dr}=-\frac{Gm(r)\rho (r)}{r^{2}}\biggl(1+\frac{P_{r}}{\rho (r)
c^{2}}\biggr)\biggl(1+\frac{4\pi r^{3}P_{r}}{m(r)c^{2}}\biggr)\biggl(1-\frac{%
2Gm(r)}{rc^{2}}\biggr)^{-1}+\frac{2\Delta }{r},
\end{equation}%
where we already defined $\Delta $ in previous section. By selecting a central energy
density $\rho _{c}$, under the boundary conditions $P_{r}(0)=P_{c}$, $m(0)=0$,
we integrate the TOV equations outwards to a radius $r=R$, at which $P_{r}$
vanishes. This yields the radius $R$ and gravitational mass $M=m(R)$ of the star.

\begin{figure}[]
\center{\includegraphics[width=8cm]{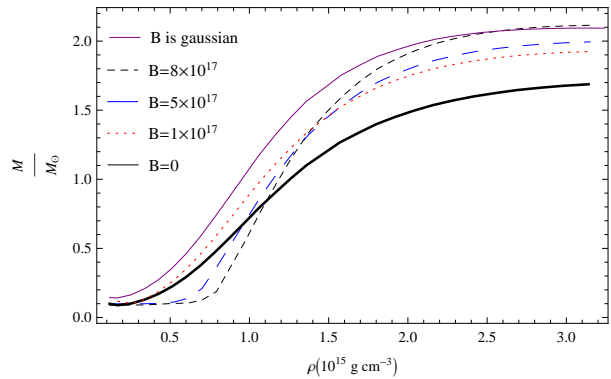}}
\caption{{\protect\small Gravitational mass vs. density for different
magnetic fields. }}
\label{me}
\end{figure}

\begin{figure}[]
\center{\includegraphics[width=8cm]{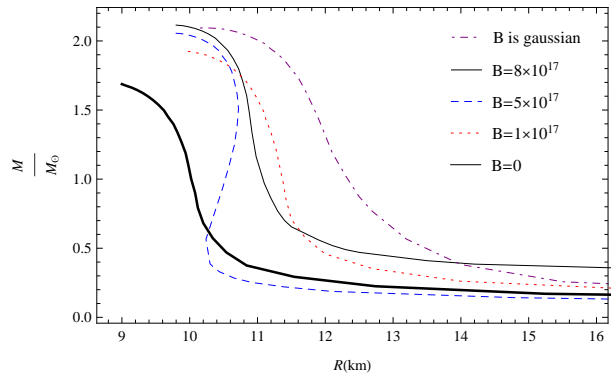}}
\caption{{\protect\small Gravitational mass vs. radius for different
magnetic fields. }}
\label{mr}
\end{figure}

The effects of magnetic fields on the gravitational mass of anisotrop neutron star
for different $B$ values are presented in Figures \ref{me} and \ref{mr} where the gravitational mass has been drawn versus the central mass density and radius ($M-R$ relation) respectively for different cases of
magnetic field. As the magnetic field increases, the gravitational mass increases.
From our results, in the absence of magnetic field, $B=0$, $M_{max}=1.68 M_{\odot}$ and $R=9 km$ have been obtained. In the presence of a uniform magnetic field, for $B=1\times10^{17}G$, $M_{max}=1.92 M_{\odot}$ and $R=9.8 km$ have been evaluated, while for $B=5\times10^{17}G$, $M_{max}=2.05 M_{\odot}$ and $R=9.8 km$ obtained. Also for $B=8\times10^{17}G$, we have obtained $M_{max}=2.11 M_{\odot}$ and $R=9.9 km$. For Gaussian magnetic field, our calculations lead to $M_{max}=2.09 M_{\odot}$, $R=10.03 km$. Also it should be noted that the above results for the masses obey the stability conditions that will be examined in the next sections.
\begin{table*}[tbp]
\caption{Properties of neutron star with different magnetic fields.}
\label{tab1}
\begin{center}
\begin{tabular}{||c|c|c|c|c|c|c|c||}
\hline\hline
$B(G)$ & ${M_{max}}\ (M_{\odot})$ & $R\ (km)$ & $R_{Sch}\ (km)$ & $\sigma (\frac{R_{Sch}}{R})$ & $z$ & $K(10^{-7}$ $m^{-2})$ & $\frac{4c^{2}R}{9G}\
(M_{\odot})$  \\ \hline\hline
$0$ & $1.68$ & $9.00$ & $4.93$ & $0.56$ & $0.51$ & $0.20$ & $2.63$
\\ \hline
$1\times 10^{17}$ & $1.92$ & $9.80$ & $5.63$ & $0.57$ & $0.53$ & $0.21$ & $2.95$
\\ \hline
$5\times 10^{17}$ & $2.05$ & $9.80$ & $6.01$ & $0.61$ & $0.61$ & $0.22$ & $2.96 $
\\ \hline
$8\times 10^{17}$ & $2.11$ & $9.90$ & $6.20$ & $0.63$ & $0.64$ & $0.24$ & $2.99$
\\ \hline
$Gaussian$ & $2.09$ & $10.03$ & $6.13$ & $0.61$ & $0.60$ & $0.22$ & $3.00$ \\ \hline\hline
\end{tabular}
\end{center}
\end{table*}

We have given some results for the properties of neutron star in Table \ref%
{tab1}. In order to make more investigation for anisotrop neutron star, we discuss
about the Schwarzschild radius, compactness, redshift, Kretschmann scalar
and Buchdahl-Bondi bound in the following sections.

\subsection{Schwarzschild Radius}

We obtain the Schwarzschild radius for the obtained masses in each magnetic
field. To find the Schwarzschild radius of neutron stars, we use the relation $R_{Sch}=%
\frac{2GM}{c^{2}}$. By obtaining the Schwarzschild Radius we have found the maximum amount of Schwarzschild Radius as $R_{Sch}=6.2 km$. This shows that our system can not be a black hole, and it is surely a neutron star because our result for the radius is greater than Schwarzschild Radius. The results indicate that by increasing the magnetic field, the Schwarzschild radius increases (see Table \ref{tab1} for more details).

\subsection{Compactness}

The compactness is a very
important quantity for neutron stars, and expresses the strength of the
surface gravitational field. One of the important quantities  which we
want to investigate is related to the compactness of a spherical object. It
can be defined as $\sigma =\frac{R_{Sch}}{R}$, which may be interpreted as
the strength of gravity. Our results from Table \ref{tab1} confirm that by increasing the magnetic
field, the compactness increases.

\subsection{Redshift}

Another known parameter of neutron stars is the gravitational redshift. The
surface gravitational redshift of a neutron star is closely connected to the
value of $\frac{M}{R}$, with $M$ being the mass and $R$ the corresponding
radius. The gravitational redshift of a neutron star is given by $z=\frac{1}{%
\sqrt{1-\frac{2GM}{c^{2}R}}}-1$. In Table \ref{tab1}, we see that the maximum value of redshift is  for $B=8\times10^{17}G$, $\sigma=0.63$. We see that these redshift values are allowed.  We can see that increasing the magnetic field
leads to increasing of the redshift.

\subsection{Kretschmann scalar}

When we study any space time, it is important above other things to know
whether the spacetime is regular or not. By regular spacetime, we simply mean
that the space time must have regular curvature invariants are finite at all
spacetime points, or contain curvature singularities at which at least one
such singularity is infinite. In the Schwarzschild metric, the components of
the Ricci tensor $(R_{\mu \nu })$ and the Ricci scalar $(R)$ are zero
outside the star, and these quantities do not give us any information about
the spacetime curvature. Therefore, we use another quantity in order to
further investigate the curvature of spacetime. The quantity that can help
us to understand the curvature of spacetime is the Riemann tensor. The
Riemann tensor may have more components, then for simplicity, we can study
the Kretschmann scalar for measurement of the curvature in a vacuum.
Therefore, the curvature at the surface of a neutron star is given as $K=%
\sqrt{R_{\mu \nu \alpha \beta }R^{\mu \nu \alpha \beta }}=\frac{4\sqrt{3}GM}{%
c^{2}R^{3}}$. The numerical results have been shown in Table \ref{tab1} where we can see
that the maximum of curvature is $8.3\times10^{-8}(m^{-2})$. Also
we have found that by increasing the magnetic field, the strength of gravity
increases.

\subsection{Buchdahl-Bondi bound}

Here, we want to investigate the upper mass limit of a static spherical
neutron star with uniform density in GR, the so-called Buchdahl theorem. The
GR compactness limit is given by $M_{BB}\leqslant \frac{4c^{2}R}{9G}$ \cite%
{Buchdahl1,Buchdahl2,Buchdahl3}, in which the upper mass limit is $M_{max}=%
\frac{4c^{2}R}{9G}$. The results of our calculations confirm that the
obtained masses of magnetic neutron stars are smaller than this limit and our system can not be a black hole.

\subsection{Dynamical Stability}

\begin{figure}[tbp]
\center{\includegraphics[width=8cm]{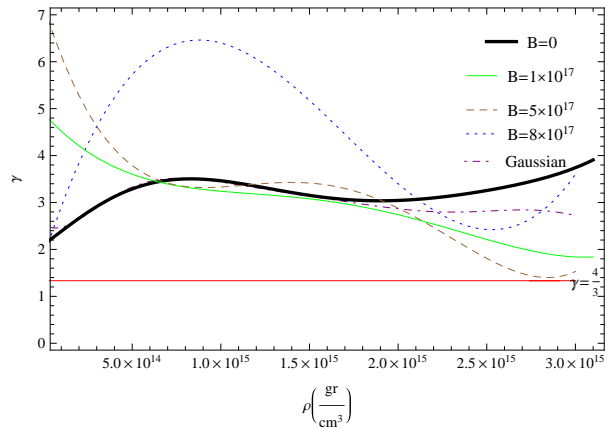}}
\caption{{\protect\small adiabatic index vs. density for different magnetic
fields. }}
\label{adia}
\end{figure}

The virial relation for an equilibrium configuration, as suggested by
Chandrasekhar and Fermi \cite{Chandrasekhar} can be written as
\begin{equation}
3(\gamma -1)\epsilon _{k}+\epsilon _{B}+\epsilon _{G}=0,  \label{chandra}
\end{equation}%
where $\epsilon _{k}$ is the total kinetic energy of the system, $\epsilon
_{B}$ is the positive magnetic energy due to magnetic field and $\epsilon
_{G}$ is the negative gravitational potential and $\gamma $ is the adiabatic
index defined as $\gamma =\frac{\rho c^{2}+P}{c^{2}P}\frac{dP}{%
d\rho }$. The total energy of the system is given by $E=\epsilon
_{k}+\epsilon _{B}+\epsilon _{G}$, using Eq.(\ref{chandra}), the total
energy can be written as
\begin{equation}
E=-\frac{3\gamma -4}{3(\gamma -1)}(|\epsilon _{G}|-\epsilon _{B}).
\end{equation}%
For stability, the necessary condition is $E<0$, or
\begin{equation}
(3\gamma -4)|\epsilon _{G}|(1-\frac{\epsilon _{B}}{|\epsilon _{G}|})>0.
\end{equation}%
To satisfy this condition, first we should have $\gamma >\frac{4}{3}$, in order to
investigate the dynamical stability of a magnetized neutron star, we plot
the adiabatic index versus the radius in Figure \ref{adia}. As one can see,
these stars enjoy interior dynamical stability. Second condition is that the
magnetic energy should not exceed the potential of gravity ($\epsilon
_{B}>|\epsilon _{G}|$), because star is no longer stable or gravitationally
unbound. Upper limit of stability is $\epsilon _{B}=|\epsilon _{G}|$. To
investigate this condition, the magnetic energy is given by
\begin{equation}
\epsilon _{B}=\frac{B^{2}R^{3}}{6},  \label{magnener}
\end{equation}%
and the gravitational potential is given by
\begin{equation}
\epsilon _{G}=-\frac{3}{4}\frac{GM^{2}}{R}.  \label{graener}
\end{equation}%
Using Eqs.(\ref{magnener}) and (\ref{graener}) into stability condition, we
can obtain the maximum value of magnetic field $B_{max}$ when the star can be
stable
\begin{equation}
B_{max}=\sqrt{\frac{9G}{2}}\frac{M}{R^{2}}.  \label{bmax}
\end{equation}%
In our work for the maximum mass $M=2.11M_{\odot }$ and radius $R=9.9km$, it is
found that $B_{max}=9.35\times 10^{17}G$. By using Eqs.(\ref{magnener}), (%
\ref{graener}) and (\ref{bmax}), we can drive the following condition for
stability
\begin{equation}
\frac{\epsilon _{B}}{|\epsilon _{G}|}=\frac{B^{2}}{B_{max}^{2}}.
\end{equation}%
The values $\frac{B}{B_{max}}$, indicate that the predicted internal
magnetic field $B$ always be lower than the magnetic field $B_{max}$ needed
for stability. Due to this and that the maximum value of our magnetic
field which was $B=8\times 10^{17}G$, the condition of stability is
satisfied.

\section{Comparison between Theory and Observations}
One of the interesting features of our calculations is related
to the comparison of the theory and its predictions with the observational data. For this purpose, we compare our results
with the empirical evidence of the neutron stars. We present some observational data for  $4U 1608-52$ \cite{Guver},  $Vela X-1$ \cite{Rawls}, $PSR J1614-2230$ \cite{Demorest},  $PSR J0348+0432 $\cite{Antoniadis} and $PSR J0740+6620$ \cite{Cromartie} in Table  \ref{tab2}.
 According to Table  \ref{tab2} we can  see that for $PSR J1614-2230$,  $PSR J0348+0432 $ and $PSR J0740+6620$, the results of the Gaussian magnetic field are more consistent with the observational results, also for  $Vela X-1$ and  $4U 1608-52$,  the results for the case where the field is uniform are more consistent for $B=8\times 10^{17}G$ and $B=5\times 10^{17}G$, respectively.

 It should be noted that obtaining the radius by observational measurement is difficult and complex. Here, according to Figure \ref{mr}, we find radius corresponding to the observational mass for different magnetic fields, and we present this comparison in Table \ref{tab3}. Because our computational mass is smaller than the observational mass for some magnetic fields, the corresponding radius is not recorded for these values.  Given the values obtained from the observations for the mass of the following objects (Table  \ref{tab2}), we can claim that the results obtained from the theory are in agreement with the observational results. These results are much more accurate and closer to the observational data when we consider the magnetic field as Gaussian.
\begin{table*}[tbp]
\caption{Comparison of our results for the mass and radius of  anisotrop neutron stars with those of observation.}
\label{tab2}
\begin{center}
\begin{tabular}{||c|c|c|c|c|c||}
\hline\hline
$Name$ &  ${M}\ (M_{\odot})$ & $R\ (km)$ & $Our\ work$ &  ${M}\ (M_{\odot})$ & $R\ (km)$\\ \hline\hline
$PSR J0740+6620$ & $2.10$ & $12(\pm2)$ & $For\ Gaussian$ & $2.09$ & $10.03$
\\ \hline
$PSR J0348+0432$ & $2.01$ & $13(\pm2)$ & $For\ B=8\times 10^{17}G$ & $2.11$ & $9.90$
\\ \hline
$PSR J1614-2230$ & $1.97$ & $12(\pm2)$ &  $For\ B=5\times 10^{17}G$ & $2.05$ & $9.80$
\\ \hline
$Vela $X-1$$ & $1.80$ & $11(\pm2)$ & $For\ B=1\times 10^{17}G$  & $1.92$ & $9.80$
\\ \hline
$4U 1608-52$ & $1.74$ & $9(\pm1)$ & $For\ B=0$ & $1.68$ & $9.00$
\\ \hline\hline

\end{tabular}
\end{center}
\end{table*}
\begin{table*}[tbp]
\caption{Radius corresponding to observational mass .}
\label{tab3}
\begin{center}
\begin{tabular}{||c|c|c|c|c|c||}
\hline\hline Observational & Observational  & Radius for  & Radius for & Radius for   & Radius for\\
  Object& mass  ${M}\ (M_{\odot})$ & $B=1\times 10^{17}G$ &  $B=5\times 10^{17}G$ &$B=8\times 10^{17}G$ &Gaussian  \\ & & $\ (km)$&$\ (km)$ &$\ (km)$ & $\ (km)$\\ \hline\hline
$PSR J0740+6620$ & \multicolumn{1}{|c|}{$2.10$}  & $-$ & $-$ & $10.00$ & $10.50$
\\ \hline
$PSR J0348+0432$ & $2.01$  & $-$ & $10.20$ & $10.40$ & $10.95$
\\ \hline
$PSR J1614-2230$ & $1.97$  & $-$ & $10.30$ & $10.55$ & $11.20$
\\ \hline
$Vela $X-1$$ & $1.80$ & $10.56$ & $10.58$ & $10.75$ & $11.55$
\\ \hline
$4U 1608-52$ & $1.74$  & $10.80$ & $10.60$ & $10.80$ & $11.60$
 \\ \hline\hline

\end{tabular}
\end{center}
\end{table*}

\section{Conclusions}

In this paper we studied the structure of anisotropic neutron stars. For this purpose, we calculated the equation of state of
magnetized neutron star that contains pure neutron matter in the presence of
an strong magnetic field. For computation of anisotropic pressure, we considered
two type of magnetic field. In the first case, $B$ is constant from the center
to the surface of the neutron star, and calculations were performed for four
fixed values of magnetic fields. In the second case, we used a Gaussian function to define
density dependent magnetic field where the maximum amount of magnetic field
is in the center of the star and decreases as it moves to the surface of the
star. Using the obtained equation of state, we integrated the
generalized TOV equation to compute the structure of anisotrop neutron star. We
showed that by increasing the magnetic field, the maximum mass of neutron
star increases. For the maximum magnetic field $B=8\times10^{17}G$, the
maximum amount of mass was obtained as $M_{max}=2.11 M_{\odot}$. We also examined
the stability of this star, where we showed that the adiabatic index is
higher than $\gamma=\frac{4}{3}$, and the star is also stable in our
computational magnetic fields. By considering different values of magnetic
fields, we evaluated compactness, redshift and the Kretschmann scalar for
these compact objects. Our results indicate that these quantities are
increasing functions of magnetic field strength.
Finally, we compared our theoretical results with the observational data,
and it was shown that our theoretical results are in agreement
with the empirical evidence of neutron stars.

\acknowledgments
We wish to thank Shiraz University Research Council.
We also wish to thank B. Eslam Panah (University of Mazandaran) for his useful comments and discussions during this work.


\end{document}